%% file: main.tex
\documentclass[runningheads]{llncs}

\usepackage[T1]{fontenc}

\usepackage{graphicx}

\usepackage{tabularx}
\usepackage[table]{xcolor}
\usepackage{booktabs}
\usepackage{threeparttable}
\usepackage{multirow, makecell}
\usepackage{xltabular}
\usepackage{hyperref}

\newcolumntype{L}[1]{>{\raggedright\arraybackslash}m{#1}}
\newcolumntype{C}[1]{>{\centering\arraybackslash}m{#1}}

\emergencystretch=\maxdimen
\begin{document}

\title{ Rethinking Self-Sovereign Identity Principles: \\An Actor-Oriented Categorization of Requirements }

\author{Daria Schumm\orcidID{0009-0004-1154-4799} \and
Burkhard Stiller\orcidID{0000-0002-7461-7463}}

\authorrunning{Schumm and Stiller}
\institute{ University of Zürich \\ 
Binzmühlestrasse 14, CH---8050 Zürich, Switzerland\\
\email{[schumm, stiller]@ifi.uzh.ch}}

\maketitle              

\begin{abstract}
    Centralized identity management systems continuously experience security and privacy challenges, motivating the exploration of Decentralized Identity (DI) and Self-Sovereign Identity (SSI) as user-focused alternatives. 
    Although prior research has consolidated SSI principles and derived quality requirements for DI/SSI systems, it is significantly limited in integrating the user viewpoint. 
    This work addresses this gap by embedding a user perspective into the requirements engineering process for DI/SSI systems. 
    Building on existing SSI principles, composite requirements were decomposed into 24 simple quality or non-functional requirements (NFR). 
    The resulting NFR are systematically mapped to the key actors, namely data owner, issuer, verifier, and system, based on varying degrees of responsibility and ownership. 
    A dependency model is introduced to formalize relationships between actors. 
    Inspired by trust modeling concepts, the model explicitly describes how actors interact and rely on each other for requirements fulfillment.
    By integrating user-centered requirements, responsibility allocation, ownership specification, and dependency modeling, this work provides the first structured model for DI/SSI system architectures. 

    \keywords{Decentralized Identity \and Self-Sovereign Identity \and Requirements Engineering \and System Modeling}
\end{abstract}


\input{sections/1_introduction}
\input{sections/2_methodology}
\input{sections/3_results}
\input{sections/4_discussion}

\input{sections/5_conclusions}






\bibliographystyle{splncs04}
\bibliography{ref}







\end{document}

%% file: sections/1_introduction.tex
\section{Introduction}

    With the growing number of vulnerabilities in centralized identity management, illustrated by numerous data breaches \cite{b,c}, other alternatives should be considered to enable more secure handling of personal data. 
    Decentralized Identity (DI) and Self-Sovereign Identity (SSI) offer an alternative to traditional identity management approaches by enabling data owners to retain control over their identity data without reliance on centralized services. 
    Existing DI/SSI systems largely build on SSI principles, such as existence, control, and persistence, originally introduced by \cite{61}.
    Other works, such as \cite{17}, extended these principles and provided more elaborate descriptions and classifications based on multiple sources (\textit{e.g.}, \cite{61,94,91,291}).
    \cite{17} outlined a comprehensive set of quality requirements that compose a DI/SSI system.
    The authors classified identified requirements into five categories, namely security, privacy, adoption and sustainability, usability and user experience, and controllability, and prioritized them based on the expert's feedback.
    The main limitation of \cite{17} is the lack of an individual user perspective on requirements, which largely misrepresents the actual users and user-focused approach of DI/SSI. 
    Similarly, DI/SSI design patterns, presented in \cite{292,294}, provide a limited perspective on user roles and requirements but comprehensively outline the functionality and interactions of DI/SSI systems.
    Design patterns are frequently used in software engineering to define how a certain objective in software development can be achieved and to describe a solution to a recurring problem \cite{295}.
    In the context of DI/SSI, there is currently no explicit mapping between the established design patterns and the quality requirements they are intended to address. 
    
    The objective of this work is to address the lack of a user viewpoint within the DI/SSI domain and to embed the user into the requirements engineering process. 
    To address this objective, this work builds on the requirements aggregated by \cite{17}.
    Some composite requirements (\textit{e.g.}, existence and representation) have been broken down into independent, simple quality requirements or non-functional requirements (NFR), resulting in a total of 24 requirements, as outlined in Table \ref{tab:nfr}.
    Design patterns, outlined in \cite{292,294}, were considered to identify goals, interests, and involvement of various users (or actors) within DI/SSI systems, as well as explicitly map them to the requirements presented in Table \ref{tab:nfr}. 
    
    As a result, this work contributes (i) a refined set of user-centered requirements that consistently reflect a user-based viewpoint, (ii) specification of responsibilities and ownership for each requirement across users, and (iii) a dependency model that specifies where, when, and by whom each requirement is realized within DI/SSI systems. 

    This work is organized as follows. 
    Section \ref{sec:definitions_methodology} specifies a generalized use case, definitions used throughout the paper, and the methodology that was used to map and categorize requirements.
    Section \ref{sec:mapping} maps design patterns and NFR, followed by Section \ref{sec:cateorization}, which provides a concise discussion of each NFR concerning its assignment. 
    Section \ref{sec:discussion} summarizes the results of NFR categorization and introduces the dependency model for DI/SSI systems.
    Section \ref{sec:conclusions} concludes the work and indicates future work. 
    
    \input{tables/nfr}

%% file: tables/nfr.tex
\begin{table*}[!htbp]
    \centering
    \caption{Non-Functional Requirements of DI and SSI Systems}
    \label{tab:nfr}
    \def\arraystretch{1.5}%
    \footnotesize
    \rowcolors{2}{white}{gray!10}
    \begin{tabularx}{\textwidth}
        {>{\hsize=0.1\hsize\linewidth=\hsize}X 
         >{\hsize=0.2\hsize\linewidth=\hsize}X 
         >{\hsize=0.7\hsize\linewidth=\hsize}X}
        \toprule
        \textbf{Key} & \textbf{Quality} & \textbf{Description} \\
        \midrule
        NFR1  & Accessibility    & User must be able to access and retrieve data \\
        NFR2  & Authenticity     & Source of identity data must be trustworthy and provable \\
        NFR3  & Autonomy         & User must be able to manage their identity independently \\
        NFR4  & Availability     & Identity data must be available at any time \\
        NFR5  & Compatibility    & Identity data must be compatible with legacy systems \\
        NFR6  & Consent          & User must explicitly consent to the use of their data \\
        NFR7  & Control          & User must be able to control access to their identity data \\
        NFR8  & Cost             & All components must have minimal costs \\
        NFR9  & Decentralization & All components should not rely on centralized elements \\
        NFR10 & Existence        & User identity must have an independent existence without relying on other services \\
        NFR11 & Interoperability & Identity data must be usable across different platforms and services \\
        NFR12 & Persistence      & Identity data must remain valid and accessible for as long as necessary \\
        NFR13 & Portability      & User must be able to move their identity data \\
        NFR14 & Privacy          & User must be able to minimize information required to share \\
        NFR15 & Protection       & Identity data must be protected against misuse \\
        NFR16 & Recoverability   & User must be able to recover identity data in case of loss and compromise \\
        NFR17 & Representation   & Users must be able to create multiple identities \\
        NFR18 & Security         & All components must ensure the data is secure \\
        NFR19 & Single Source    & User must be the single authoritative source of their identity \\
        NFR20 & Standard         & Credentials must adhere to open standards \\
        NFR21 & Transparency     & Information about data use must be readily available \\
        NFR22 & Usability        & User must be able to use their data efficiently and intuitively \\
        NFR23 & User Experience  & Identity management process must be simple, consistent, and user-friendly \\
        NFR24 & Verifiability    & Identity data must be verifiable \\
        \bottomrule
    \end{tabularx}
\end{table*}

%% file: sections/2_methodology.tex
\section{Use Case, Definitions, \& Methodology}\label{sec:definitions_methodology}
    
    Categorizing requirements requires a clear understanding of who is responsible for what within the DI/SSI system. 
    First, a generalized use case is described to illustrate the fundamental processes common to all DI/SSI systems. 
    Then, the users of a DI/SSI system are defined, along with key concepts, such as responsibility and ownership. 
    Lastly, the three-step methodology for systematic requirements categorization is outlined. 

    \subsection{Generalized Use Case}
        Use cases provide a structured mechanism for analyzing responsibilities and requirements within a system \cite{334}.
        The DI/SSI domain has numerous use cases, ranging from healthcare to education. 
        By omitting implementation details, each use case reveals a recurring pattern.
        For example, a DI/SSI system for a government service and for event ticketing might have different goals, but they represent the same underlying functionalities and involve the same core stakeholders or users (\textit{e.g.}, credential issuer and verifier), and rely on the same fundamental processes (\textit{e.g.}, credential issuance and verification). 
        A shared goal of both systems is the ``identification of a human”, adapted to a different context.

        The validity of a generalized use case argument is supported by evidence from existing systems.
        For example, Sovrin \cite{111} and IDunion \cite{218} operate in different regulatory and operational environments. 
        However, they both rely on common architectural components (\textit{e.g.}, a data owner) and technical features (\textit{e.g.}, Verifiable Credentials (VC)) to achieve their goals. 
        Furthermore, the systems share comparable design choices in credential lifecycle management and privacy-preserving user control, including credential issuance, presentation, and verification. 
        These recurring patterns across implementations suggest a common underlying model that captures these fundamental processes. 
        Moreover, the existence of design patterns for DI/SSI systems further suggests that certain aspects are not only generalizable but are necessary to describe the core functionality and capabilities of any DI/SSI system.  

        To formulate a generalized use case, this work follows the approach outlined in \cite{334}. 
        The method begins by defining the final state of the system and then expands it into intermediary steps required to reach that state. 
        In a DI/SSI system, the end goal is for a data owner to get authenticated and authorized for a specific service of their choice. 
        The use case, therefore, begins with the data owner requesting a credential from an issuer. 
        The issuer then issues a credential, either using an existing credential schema or creating a new one, based on the information it holds about the data owner. 
        As the next step, the data owner presents a credential to the verifier for verification. 
        During verification, the verifier verifies the issuer's signature and checks whether the credential has been revoked. 
        Once verification is complete, the verifier considers the data owner authorized to access the service or authenticated and enables access. 

    \subsection{Definitions}
        Each requirement outlined in Table \ref{tab:nfr} was assigned to one or more relevant primary users, namely data owner, issuer, or verifier, as defined by the W3C standards \cite{23,58,161}.
        A system provides an interface to users and is treated as a separate component in this work because it supports multiple user interactions. 
        A system is divided into global and local components, indicating that it has two interfaces.
        The following definitions of components are used throughout this work: 
        \begin{itemize}
            \item The \textit{data owner} ($o$; or identity owner, DID subject) is the entity that receives, stores, and presents a credential (or VC). 
            \item The \textit{issuer} ($i$) creates and signs credentials, asserting claims about the data owner.
            \item The \textit{verifier} ($v$) requests and verifies the validity of credentials presented by the data owner. 
            \item A \textit{global system} ($s$), referred to simply as a system, represents a blockchain component that enables decentralization of the identity management system.
            \item A \textit{local system}, referred to as a wallet ($w$), is software that the data owner interacts with. 
        \end{itemize}
        
        Each component can be referred to as an \textit{actor}.
        Actors are domain stakeholders who interact with one another to achieve goals \cite{323}. 
        An actor has a goal within a system and represents agents, such as an organization or a person \cite{320}. 
        A \textit{goal} is the interest of an actor, which can be satisfied with a specific task \cite{320}.
        
        An actor has a certain level of responsibility for an NFR.
        The \textit{responsibility} is a relationship between two actors with respect to a particular state of affairs, capturing the underlying reasons why an actor performs a given action \cite{317}. 
        Responsibilities can be satisfied \textit{directly} by performing a role or \textit{indirectly} by delegating responsibility to another actor \cite{317}. 
        A \textit{role} refers to “what has to be done” and provides an abstraction for actions, which change the system state (in other words, functions) \cite{317}.
        In this work, responsibility is classified as primary (direct), secondary (supporting), or tertiary (indirect influence). 
        \begin{itemize}
            \item \textit{Primary} responsibility refers to the fundamental responsibility that an actor performs directly and guarantees its fulfillment. 
            \item \textit{Secondary} responsibility reflects a supporting role, where an actor facilitates the performance of another actor’s responsibility. 
            \item \textit{Tertiary} responsibility indicates indirect responsibility, where an actor benefits from or relies on others to perform the fulfillment. 
        \end{itemize}
        
        Similarly, an actor may have \textit{ownership} over NFR. 
        Ownership represents which actor is an owner of a service, goal, task, or resource \cite{318,320}.
        The owner has exclusive authority over a service or access, and their enforcement \cite{320}. 
        
    \subsection{Methodology}
        Each assignment of a requirement to an actor is supported by a concise justification, based on actor roles, responsibilities, and architectural relevance.
        The justification was formulated by considering whether an actor has a stake in, derives benefits from, or is responsible for enforcing a particular software quality, drawing on relevant literature, particularly \cite{17,292,294}. 
        Overall, the categorization process followed three steps methodology, namely (i) identification of goals and interests of actors, (ii) responsibility, and (iii) ownership assignments. 
        
        As the first step, the goals and interests of each actor were identified through a literature review of DI/SSI design patterns (\textit{e.g.}, \cite{17,292,294}) and relevant standards (\textit{e.g.}, \cite{23,58,161}). 
        The analysis focused on identifying which software qualities each actor is inherently responsible for, benefits from, or relies on. 
        For example, privacy preservation for the data owner, as the data is owned by them and they have a primary interest in keeping it private. 
        Similarly, verifiability for the verifier, because the primary goal of the verifier is to verify a credential. 
        As the second step, the relevance and impact of each NFR on actors were evaluated through a structured assessment that assigned each actor a degree of responsibility for the NFR.
        For example, the data owner is responsible for storing their own data, thereby bearing primary responsibility for accessibility. 
        Lastly, ownership was assigned to each actor. 
        For example, the data owner owns their personal data, whereas the verifier should seek consent to process it. 
        

%% file: sections/3_results.tex
\section{Mapping Design Patterns to Requirements}\label{sec:mapping}

    This section maps DI/SSI requirements to the design patterns described in \cite{292} and \cite{294}, which serve as the basis for the requirements categorization presented in the next section. 
    Although design patterns are not requirements specifications, they still implicitly provide necessary information about functional requirements (FR) and NFR. 
    For example, \cite{294} outlines the benefits of the ``Master and Sub Key Generation” pattern as identifiability, privacy, and availability.
    In addition to these qualities, the pattern also implies functional aspects, such as ``master-key manages sub-keys”, ``sub-keys used for signing messages”, and ``sub-key is linked to a unique identifier”. 
    These statements are similar to FR, as they describe specific system capabilities and behaviors. 
    However, not all identified benefits correspond directly to the NFR used in this work. 
    For instance, ``identifiability”, mentioned in \cite{294}, does not directly match any of the NFR in Table \ref{tab:nfr}, while other benefits, such as ``privacy” and ``availability”, align well with NFR. 
    
    Both \cite{292} and \cite{294} specify the quality benefits of each design pattern. 
    Therefore, it is possible to systematically map these patterns to NFR they intend to support.
    Table \ref{tab:mapping_example} presents an example of the resulting mapping between five NFR and their corresponding design patterns, based on the benefits described.
    Several NFR identified in this work are not addressed by the existing design patterns. 
    In particular, no design pattern explicitly supports the autonomy, compatibility, consent, existence, persistence, and user experience requirements.
    Despite these gaps in design patterns, the mapping provides a valuable starting point for systematic analysis of NFR. 
    The complete mapping between design patterns and NFR is provided in Table 2 of the Appendix\footnote{https://github.com/schummd/categorization}.

\input{tables/mapping_example}

\section{Categorizing Requirements}\label{sec:cateorization}

    Following the methodology outlined previously, 24 NFR were analyzed and categorized, resulting in the assignment of each NFR to an actor, the justification for the assignment, as well as the identification of responsibility and ownership of an actor over NFR. 
    Table \ref{tab:categorization} summarizes the categorization results.

\input{tables/categorization}

    \paragraph{NFR1:}
        Accessibility may refer to two different things. 
        First, accessibility ensures that the data is in an accessible format, allowing users, regardless of ability, to use it \cite{161}. 
        The typical approach to ensure accessibility is to follow a certain standard on data representation and format, such as DID or VC, defined by W3C \cite{58,161}
        However, in DI/SSI, accessibility refers to data accessibility. 
        That is, how easy it is to locate and access the data. 
        The data owner is responsible for storing their own data and, thus, is the primary actor responsible for locating and accessing it. 
        The verifier has a tertiary responsibility because they benefit from the data owner performing the fulfillment. 
        In other words, the verifier uses the data provided by the data owner. 
        The issuer has no direct responsibility for data access. 
        Nevertheless, they need to access credential schemas to enable effective credential issuance \cite{84}.
        Therefore, an issuer is not responsible for the definition of accessibility used in this work. 
        Ownership of NFR1 is assigned to the data owner, who is primarily responsible for storing, accessing, and retrieving the data, and to the local system (wallet), which enforces accessibility.
    
    \paragraph{NFR2:}
        The issuer is primarily responsible for the authenticity of the credential. 
        Authenticity is achieved by attaching a signature to the credential, which can later be verified to attest to the authenticity of the information \cite{17}.
        The verifier verifies the signature created by the verifier (the responsibility of the issuer), thereby fulfilling secondary responsibility. 
        The data owner has a tertiary responsibility because they benefit from the issuer's fulfillment of that responsibility. 
        That is, an authentic and verifiable signature ensures the credential is valid and the data owner can access or use a service. 
        Ownership of NFR2 is assigned to the issuer, who is primarily responsible for the correctness and authenticity of the issued credentials. 
    
    \paragraph{NFR3:}
        Autonomy NFR refers to the ability of the data owner to manage their identity independently of other actors and entities. 
        The data owner is primarily responsible for this NFR and is the owner, while the verifier and issuer are not involved in this NFR. 
    
    \paragraph{NFR4:}
        Availability refers to the continuous availability of identity data. 
        Since the data owner is responsible for storing and retrieving the data, they must also ensure that the data is available to them at any time and others when necessary \cite{17}. 
        Thus, the data owner owns and is primarily responsible for NFR4. 
        The ownership of NFR also belongs to the local system (Wallet) because it supports its technical enforcement.
    
    \paragraph{NFR5:}
        The identity data must be compatible with legacy systems, other wallets, and infrastructures \cite{17}. 
        Since the issuer is primarily responsible for issuing a credential and ensuring it adheres to the standards, the issuer is also responsible for and owns the compatibility NFR. 
        The verifier has a secondary responsibility because it facilitates verification, which may be performed by a different system than the one in which the credential was issued.
        Thus, verifiers facilitate the performance of compatibility. 
        The data owner has the tertiary responsibility over compatibility, since they benefit from credentials being widely usable and verifiable across different infrastructures. 
        However, it is important to note that compatibility is not visible to the data owner due to wallet abstractions \cite{17}.  
    
    \paragraph{NFR6:}
        The data owner has primary responsibility for providing consent to any access to and use of their personal data \cite{17,84}.
        Additionally, the verifier is primarily responsible for obtaining the data owner's consent for the collection, processing, and storage of data \cite{17}. 
        Moreover, the issuer is primarily responsible for obtaining the data owner's consent before issuing a credential that contains personal data \cite{17}. 
        However, this is typically done outside of the DI/SSI system.
        Ownership of NFR6 lies with the data owner. 
    
    \paragraph{NFR7:}
        Control refers to a user's ability to manage access to their personal data. 
        Since the data owner is responsible for storing and retrieving the data, control lies with them. 
        Neither the verifier nor the issuer has control over the data owner's personal data. 
        A local system or a wallet also has ownership of NFR7 because it technically supports data protection.
    
    \paragraph{NFR8:}
        Cost affects all actors of a DI/SSI system. 
        Data owner, issuer, and verifier all have a tertiary responsibility, since there is no direct fulfillment responsibility, but they all benefit from its fulfillment.  
        The ownership of this NFR lies with the DI/SSI (global) system because the costs are rooted in the software application's design choices. 
    
    \paragraph{NFR9:}
        Similar to cost (NFR8), decentralization affects the main actors but is not their responsibility.
        Data owner, issuer, and verifier all have a tertiary responsibility, as they benefit from it but are not responsible for its fulfillment. 
        The ownership belongs to the DI/SSI (global) system. 
    
    \paragraph{NFR10:}
        The existence means that the data owner’s identity has an independent existence and does not rely on other services \cite{17}. 
        This NFR is indirectly fulfilled by the data owner and facilitates the performance of other responsibilities, such as accessibility (NFR1), availability (NFR4), and control (NFR7). 
        Therefore, the data owner has a primary responsibility over NFR10. 
        Both verifier and issuer should also exist independently, but do not directly interact with this NFR. 
        Ownership of NFR10 lies with the data owner. 
    
    \paragraph{NFR11:}
        Interoperability is closely related to compatibility (NFR5) and standard (NFR20).
        Similar to compatibility (NFR5), the issuer is primarily responsible for issuing credentials in an interoperable format, while the verifier has a secondary responsibility.
        The data owner has a tertiary responsibility since they benefit from the fulfillment of interoperability by other actors. 
        The ownership of NFR11 belongs to the issuer.
    
    \paragraph{NFR12:}
        Persistence aligns with accessibility (NFR1) and availability (NFR4) by ensuring that data is persistent and accessible over time. 
        Since the data owner is responsible for storing and accessing the data, they are also primarily responsible for retaining it for as long as necessary. 
        Ownership of this NFR lies with the data owner. 
    
    \paragraph{NFR13:}
        The data owner should be able to move their personal data from one system to another \cite{84}. 
        This makes the data owner beneficiary of the fulfillment of portability, but not responsible for its fulfillment. 
        Thus, the data owner has a tertiary responsibility, and ownership lies with a wallet (local system). 
        The wallet enables the export and import of identity data and credentials. 
    
    \paragraph{NFR14:}
        Privacy in the DI/SSI domain refers to data minimization. 
        The data owner should be able to select a minimal set of personal data to be shared. 
        This implies that the data owner is primarily responsible for maintaining the privacy of their data, as they make the decisions about sharing the information. 
        Importantly, the verifier also shares primary responsibility for privacy, as minimal disclosure would not be achieved if the verifier requests more information than necessary. 
        The issuer has a secondary responsibility since they issue a credential that supports selective disclosure. 
        Ownership of NFR14 is shared among the data owner, verifier, and issuer. 
    
    \paragraph{NFR15:}
        According to the protection NFR, personal data should be protected against misuse. 
        Since the data owner is primarily responsible for storing and managing his personal data, protection also falls within his primary responsibility.
        This aligns with accessibility (NFR1), availability (NFR4), control (NFR7), existence (NFR10), and representation (NFR17).  
        The data owner is responsible for protecting data during storage and transmission. 
        In DI/SSI systems, the verifier typically does not store personal data in plain text. 
        Thus, the verifier shares no responsibility for data protection. 
        However, the verifier may request more data than the minimum subset, thereby making it primarily responsible for data protection. 
        Lastly, the issuer shares some responsibility for data protection, but not within the DI/SSI system. 
        The issuer holds data about the data owner before issuing a credential (\textit{e.g.}, a university has records of a student before issuing a digital certificate attesting to their successful graduation). 
        This does not fall under the NFR of DI/SSI systems and, as a result, the issuer shares no responsibility in NFR15. 
        Ownership of this NFR lies with the data owner and the local system (wallet). 
    
    \paragraph{NFR16:}
        The data owner has primary responsibility for data recoverability, as they store and manage it, aligning with accessibility (NFR1), availability (NFR4), control (NFR7), existence (NFR10), and protection (NFR15).
        There are multiple approaches to recoverability depending on how the data was lost (\textit{e.g.}, compromise, physical loss of a mobile phone). 
        Therefore, there are multiple ways to fulfill recoverability, some of which depend on the DI/SSI system capabilities (\textit{e.g.}, a wallet provides a backup). 
        Ownership of this NFR is held by the wallet (local system). 
    
    \paragraph{NFR17:}
        Representation refers to the data owner's ability to create multiple identities for different interactions \cite{17}. 
        Therefore, the data owner has primary responsibility for and ownership of NFR17. 
    
    \paragraph{NFR18:}
        Security has multiple definitions, and within the DI/SSI domain, it refers to the general principle that data should be secure at all times. 
        \cite{17} points out that the data owner should have confidence in the software and other actors of the system, as well as overall security associated with data storage and transmission. 
        This implies that security is a service owned by the DI/SSI (global) system rather than by specific actors who use it, while the wallet (local system) enforces security features to protect data storage. 
        Therefore, the data owner benefits from the security but does not always enforce it or guarantee its fulfillment, indicating that the data owner has tertiary responsibility. 
        Similar to the data owner perspective and protection (NFR15), the data should remain secure during verification and credential issuance. 
        However, this security should be ensured by the inherent software design, while the verifier and issuer benefit from its fulfillment. 
        Thus, the verifier and issuer have tertiary responsibility. 
    
    \paragraph{NFR19:}
        The data holder should be the single authoritative source of their personal identity data \cite{17}. 
        Thus, the data owner has primary responsibility for and ownership of this NFR. 
        The verifier and issuer do not share any responsibility nor benefit from it. 
    
    \paragraph{NFR20:}
        The standard closely aligns with compatibility (NFR5) and interoperability (NFR11), explicitly stating that credentials should follow a specific open standard \cite{17}. 
        The issuer is primarily responsible for guaranteeing the fulfillment of this NFR.  
        While adherence to the standards is not explicit to the data holder, they benefit from the fulfillment of this NFR because it provides more flexibility and facilitates interoperability (NFR11), implying tertiary responsibility. 
        The verifier has a secondary responsibility because they can specify which credential standards they can verify.
        As a result, the issuer is fulfilling its responsibility to adhere to a specified standard. 
        The ownership of this NFR belongs to the issuer. 
    
    \paragraph{NFR21:}
        Transparency in the context of DI/SSI systems refers to the readiness of the information about data use \cite{17}. 
        Since the verifier is the only actor who may request data from the data holder and use it for authorization, it bears primary responsibility for providing transparency to the data owner. 
        Moreover, \cite{84} points out that issuers should also support transparency by ensuring the credential issuance process is auditable and understandable.
        Similarly, the issuer shares the primary responsibility. 
        The data owner benefits from the fulfillment of transparency, thus sharing the tertiary responsibility. 
        The ownership of this NFR is shared between the verifier and issuer. 
    
    \paragraph{NFR22:}
        Usability ensures that the data owner can use the system and its functionality effectively and intuitively \cite{17}. 
        While a data holder benefits from the system's usability, they are not responsible for its fulfillment, indicating a tertiary responsibility. 
        Since the verifier and issuer do not have a usability requirement, they are not responsible for it. 
        Ownership of this NFR is held by the wallet (local system). 
    
    \paragraph{NFR23:}
        While usability (NFR22) and this NFR are closely related, user experience refers to the support of identity management processes being simple, consistent, and user-friendly \cite{17}. 
        Therefore, similarly to usability (NFR22), the data owner benefits from a good user experience but is not directly responsible, indicating tertiary responsibility. 
        Ownership of this NFR is held by the wallet (local system).
    
    \paragraph{NFR24:}
        The verifier should be able to independently verify credentials provided by a data owner, indicating that the verifier is primarily responsible for fulfilling this NFR.
        The issuer plays a supporting role in ensuring that the credential is technically verifiable (\textit{e.g.}, it includes the necessary cryptographic elements, such as the issuer’s signature), thereby assuming a secondary responsibility. 
        The data owner benefits from the fulfillment of the NFR and relies on the verifier to fulfill it, indicating tertiary responsibility. 
        The ownership of this NFR belongs to the verifier and issuer. 
    

%% file: tables/mapping_example.tex
\begin{table*}[h]
    \centering
    \caption{Example of Mapping Design Patterns and Non-Functional Requirements}
    \label{tab:mapping_example}
    \def\arraystretch{1.5}%
    \footnotesize
    \rowcolors{2}{gray!10}{white}
    \begin{tabularx}{\textwidth}
        {>{\hsize=0.1\hsize\linewidth=\hsize}X 
         >{\hsize=0.2\hsize\linewidth=\hsize}X 
         >{\hsize=0.35\hsize\linewidth=\hsize}X
         >{\hsize=0.35\hsize\linewidth=\hsize}X
         }
        \toprule
        \rowcolor{white}
        \textbf{Key} & \textbf{NFR} & \textbf{Design Patterns \cite{292}} & \textbf{Design Patterns \cite{294}} \\
        \midrule
        
        NFR1  & Accessibility & Public Institution Registry; Trusted Schemas Registry; Status Registry; Decentralized Identifier (DID) Registry; Public DIDs; Local (Private) Storage; External (Remote) Cloud Storage & - \\

        NFR2  & Authenticity & Verifiable ID; Dual Resolution; Qualified Verifiable Credentials (VC; VC signing); Binding VCs and Qualified Electronic Certificates & - \\

        NFR3  & Autonomy & - & - \\

        NFR4  & Availability & Status Registry; DID Registry; Public DIDs & Master and Sub Key Generation; Multiple Registration \\

        NFR5  & Compatibility & - & - \\
        
        \bottomrule
    \end{tabularx}
\end{table*}

%% file: tables/categorization.tex
\begin{table*}[]
    \centering
    \caption{ Categorization, Responsibility, and Ownership of Non-Functional Requirements }
    \label{tab:categorization}
    \def\arraystretch{1.5}%
    \footnotesize
    \rowcolors{2}{white}{gray!10}
    \begin{tabularx}{\textwidth}
        {>{\hsize=0.1\hsize\linewidth=\hsize}X 
         >{\hsize=0.25\hsize\linewidth=\hsize}X 
         >{\hsize=0.20\hsize\linewidth=\hsize}X
         >{\hsize=0.20\hsize\linewidth=\hsize}X
         >{\hsize=0.20\hsize\linewidth=\hsize}X
         >{\hsize=0.20\hsize\linewidth=\hsize}X}
        \toprule
        \rowcolor{white} & & \multicolumn{3}{c}{\textbf{Responsibility}} & \\ 
        \rowcolor{white} \textbf{Key} & \textbf{Quality} & \textbf{$o$} & \textbf{$v$} & \textbf{$i$} & \textbf{Ownership}\\
        \midrule
        NFR1  & Accessibility    & Primary   & Tertiary  & -         & $o$, $w$ \\
        NFR2  & Authenticity     & Tertiary  & Secondary & Primary   & $i$      \\
        NFR3  & Autonomy         & Primary   & -         & -         & $o$      \\
        NFR4  & Availability     & Primary   & -         & -         & $o$, $w$ \\
        NFR5  & Compatibility    & Tertiary  & Secondary & Primary   & $i$      \\
        NFR6  & Consent          & Primary   & Primary   & Primary   & $o$      \\
        NFR7  & Control          & Primary   & -         & -         & $o$, $w$ \\
        NFR8  & Cost             & Tertiary  & Tertiary  & Tertiary  & $s$      \\
        NFR9  & Decentralization & Tertiary  & Tertiary  & Tertiary  & $s$      \\
        NFR10 & Existence        & Primary   & -         & -         & $o$      \\
        NFR11 & Interoperability & Tertiary  & Secondary & Primary   & $i$      \\
        NFR12 & Persistence      & Primary   & -         & -         & $o$      \\
        NFR13 & Portability      & Tertiary  & -         & -         & $w$      \\
        NFR14 & Privacy          & Primary   & Primary   & Secondary & $o$, $v$, $i$ \\
        NFR15 & Protection       & Primary   & -         & -         & $o$, $w$ \\
        NFR16 & Recoverability   & Primary   & -         & -         & $w$      \\
        NFR17 & Representation   & Primary   & -         & -         & $o$      \\
        NFR18 & Security         & Tertiary  & Tertiary  & Tertiary  & $s$      \\
        NFR19 & Single Source    & Primary   & -         & -         & $o$      \\
        NFR20 & Standard         & Tertiary  & Secondary & Primary   & $i$      \\
        NFR21 & Transparency     & Tertiary  & Primary   & Primary   & $i$, $v$ \\
        NFR22 & Usability        & Tertiary  & -         & -         & $w$      \\
        NFR23 & User Experience  & Tertiary  & -         & -         & $w$      \\
        NFR24 & Verifiability    & Tertiary  & Primary   & Secondary & $v$, $i$ \\
        \bottomrule
    \end{tabularx}
\end{table*}

%% file: sections/4_discussion.tex
\section{Discussion \& Dependency Model}\label{sec:discussion}

\begin{figure}[t]
        \centering
        \includegraphics[keepaspectratio=true, width=0.6\linewidth]{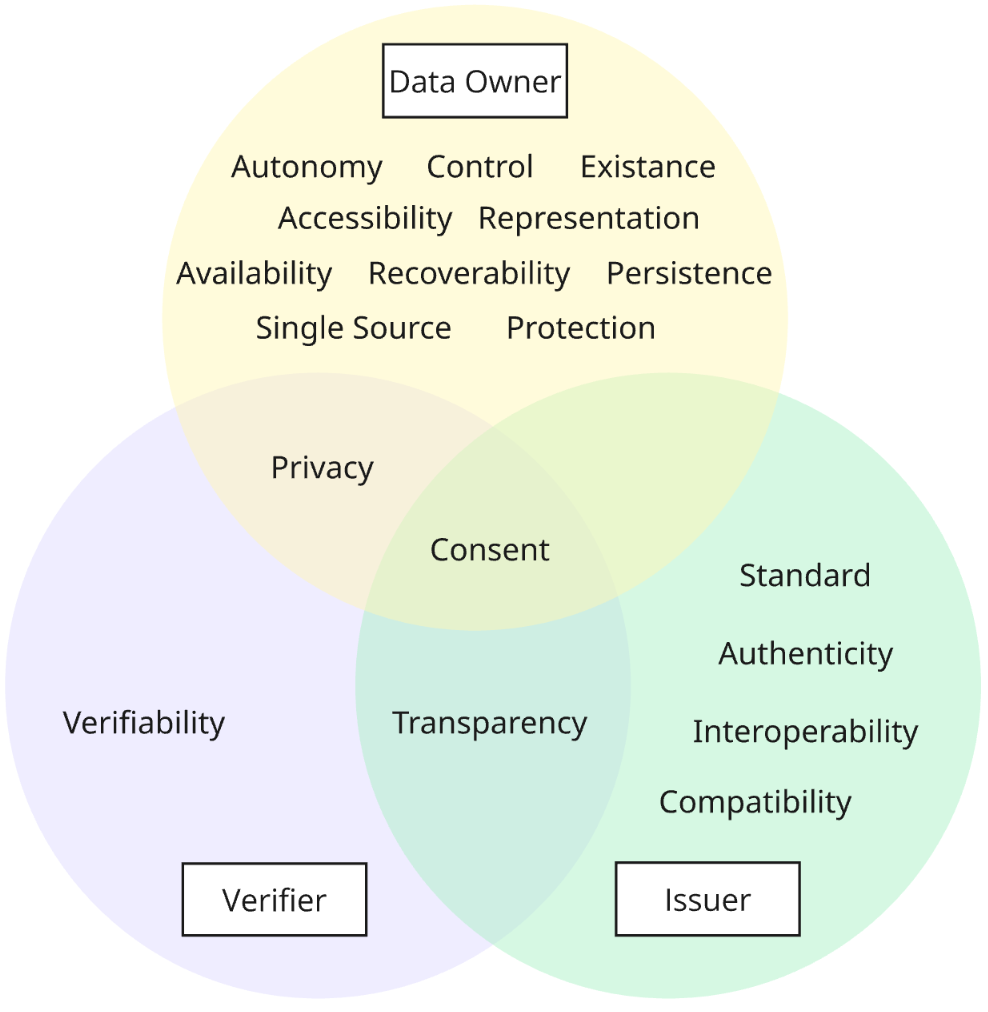}
        \caption{Primary Responsibilities Over Non-Functional Requirements}
        \label{fig:responsibility_primary}
    \end{figure}

    Based on the categorization outlined in previous section, the data owner is the most frequent actor with primary responsibility, with 11 out of 24 NFR (around 46\%) in total, namely accessibility (NFR1), autonomy (NFR3), availability (NFR4), consent (NFR6), control (NFR7), persistence (NFR12), privacy (NFR14), protection (NFR15), recoverability (NFR16), representation (NFR17), and single source (NFR19).
    This list of primary responsibilities encompasses most of the original SSI principles outlined by \cite{61}, with approximately 60\% of the original principles falling under the data owner's primary responsibility.
    This shows the importance and centrality of the data owner to a DI/SSI system. 
    The issuer is primarily responsible for six NFR (25\%), namely authenticity (NFR2), compatibility (NFR5), consent (NFR6), interoperability (NFR11), standard (NFR20), and transparency (NFR21). 
    This reflects the issuer's role in generating trusted and verifiable credentials and adhering to standards and legal frameworks. 
    The verifier is primarily responsible for five NFR (around 21\%), namely consent (NFR6), privacy (NFR14), protection (NFR15), transparency (NFR21), and verifiability (NFR24). 
    This indicates the verifier’s role in supporting privacy and protecting the data owner's data, as well as ensuring that credentials can be verified. 
    Figure \ref{fig:responsibility_primary} illustrates the primary responsibilities of each actor within the DI/SSI system. 

    The ownership of each NFR was assigned to actors according to whether the actor can provide or support the service. 
    For instance, the issuer is the owner of the authenticity (NFR2) because only the issuer can support this service. 
    In other words, authenticity originates from the issuer via cryptographic signatures. 
    Few NFR, such as cost (NFR8), decentralization (NFR9), portability (NFR13), security (NFR18), usability (NFR22), and user experience (NFR23), were assigned to the system ownership. 
    The system is not an actor or component exclusive to a DI/SSI system and is present in any software application. 
    However, within the DI/SSI system, primary actors (data owner, issuer, and verifier) are not the owners of these NFR. 
    For example, the data owner benefits from portability but is not the owner of that NFR because they share no primary responsibility in its fulfillment.
    However, the data owner relies on the system to enforce the functionality, while the system provides portability as a service to the data owner. 
    Therefore, these NFR can be treated as constraints on DI/SSI systems rather than requirements. 
    A \textit{constraint} is a requirement that ``does not add any capability to a system”, but controls how capabilities are achieved \cite{332}.
    
    Building on the results of the NFR assignment to actors, a dependency model is developed.
    The dependency model, inspired by the trust model introduced in Tropos methodology \cite{319}, represents relationships between actors and dependencies within a DI/SSI system. 
    Since DI/SSI systems typically operate in a trustless environment, supported by a neutral blockchain component, the original notion of the trust model was shifted to a dependency model to more accurately reflect the nature of the relationships between actors. 
    Even though actors within DI/SSI systems do not require mutual trust, dependencies on the fulfillment of functional requirements remain, as evident from the discussion in Section \ref{sec:cateorization}. 
    
\begin{figure*}[t]
        \centering
        \includegraphics[keepaspectratio=true, width=\linewidth]{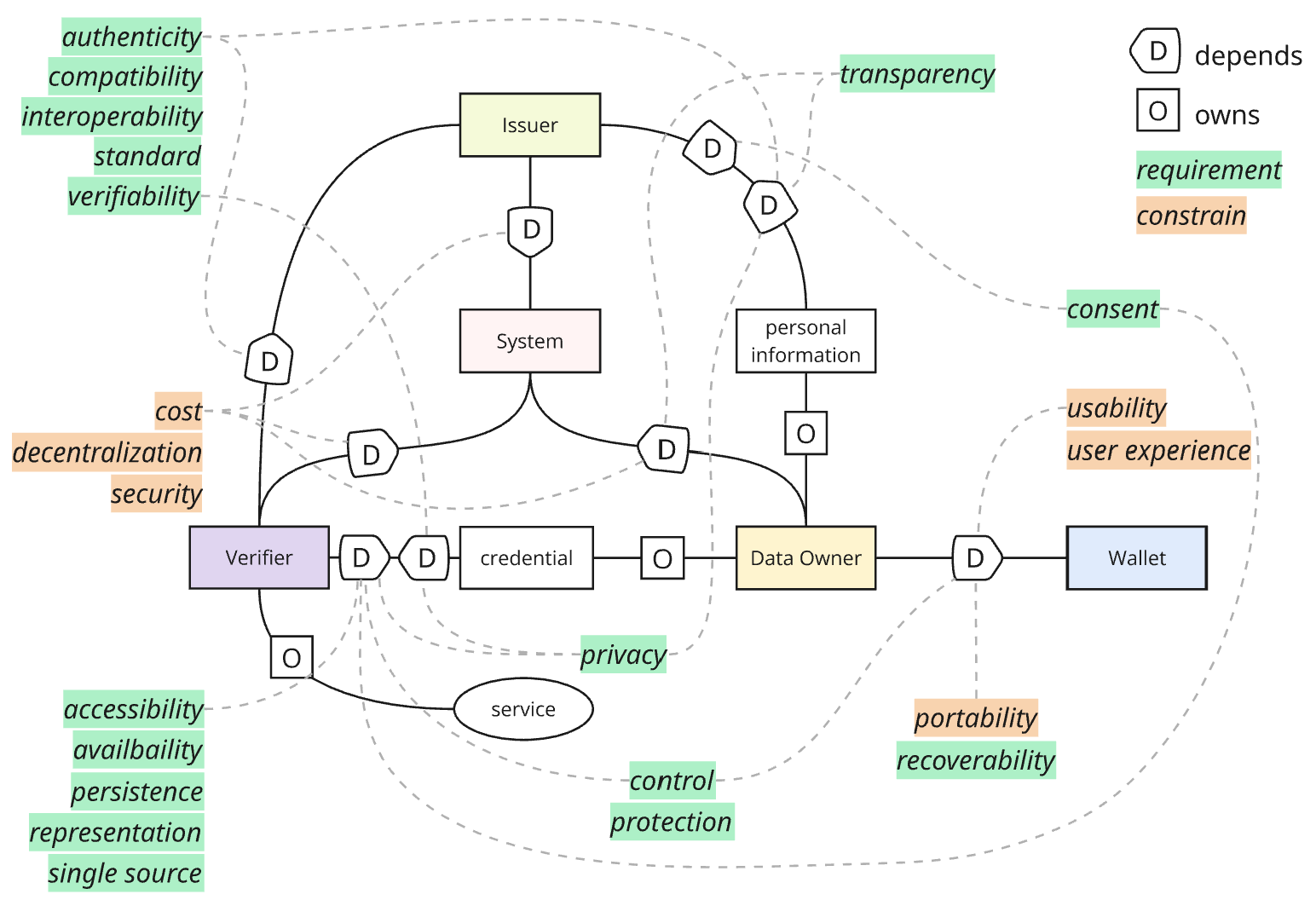}
        \caption{Dependency Model}
        \label{fig:dependency_model}
    \end{figure*}

    Particularly, the dependency model formalizes dependencies within the system, following the depends(A, B, NFR X) pattern, meaning that actor A relies on actor B to fulfill NFR X to achieve a goal of actor A. 
    The dependency model assumes that primary responsibility for NFR requires a certain degree of trust from other actors.  
    In other words, actor A must rely on actor B to fulfill NFR X. 
    For example, in the case of authenticity (NFR2), the issuer is primarily responsible for ensuring the credential is authentic. 
    At the same time, the data owner relies on the issuer to fulfill NFR2 so that they can benefit from the fulfillment (tertiary responsibility) and obtain access to a service that requires this credential. 
    The verifier does not benefit from the fulfillment of NFR2, but by performing the verification process, they provide a supporting role (secondary responsibility). 
    Therefore, to represent dependency relationships, (i) the data owner trusts the issuer with credential validity, and (ii) the verifier trusts the issuer with an authentic signature on the credential (in this instance, trust is supported by the cryptographic validity of a signature). 
    Based on this approach, the dependency relationships can be modeled, as shown in Figure \ref{fig:dependency_model}, where dependency is represented with D.
    In this model, ownership represents authority over the service or data and is denoted by O. 
    Table \ref{tab:dependencies_example} provides an example of the dependency model for five NFR. 
    The complete dependency model is available in Table 4 of the Appendix\footnote{https://github.com/schummd/categorization}. 
    
    \input{tables/dependencies_example}

%% file: tables/dependencies_example.tex
\begin{table*}[]
    \centering
    \caption{Dependencies Between Actors for Each Non-Functional Requirement}
    \label{tab:dependencies_example}
    \def\arraystretch{1.5}%
    \footnotesize
    \begin{threeparttable}
        \begin{tabularx}{\textwidth}
            {>{\hsize=0.1\hsize\linewidth=\hsize}X 
             >{\hsize=0.2\hsize\linewidth=\hsize}X 
             >{\hsize=0.05\hsize\linewidth=\hsize}X
             >{\hsize=0.05\hsize\linewidth=\hsize}X
             >{\hsize=0.3\hsize\linewidth=\hsize}X
             >{\hsize=0.3\hsize\linewidth=\hsize}X}
            \toprule
            \textbf{Key} & \textbf{NFR} & \textbf{PR\tnote{a}} & \textbf{O\tnote{b}} & \textbf{Dependency} & \textbf{Pattern} \\
            \midrule

            \rowcolor{gray!10}
            NFR1 & Accessibility & $o$ & $o$ & Verifier relies on the data owner to present accessible credential data. & depends($v$, $o$, NFR1) \\
            \rowcolor{gray!10}
            & & $w$ & $o$ & The data owner depends on the wallet to provide access to data. & depends($o$, $w$, NFR1) \\

            NFR2 & Authenticity & $i$ & $i$ & The data owner relies on the validity of credentials. & depends($o$, $i$, NFR2) \\
            & & $i$ & $i$ & Verifier relies on a valid signature of a credential. & depends($v$, $i$, NFR2) \\

            \rowcolor{gray!10}
            NFR3 & Autonomy & $o$ & $o$ & -- & -- \\

            NFR4 & Availability & $o$ & $o$ & Verifier depends on the data owner to provide credentials. & depends($v$, $o$, NFR4) \\
            & & $w$ & $o$ & The data owner depends on the wallet for the data to be accessible. & depends($o$, $w$, NFR4) \\

            \rowcolor{gray!10}
            NFR5 & Compatibility & $i$ & $i$ & Verifier relies on the issuer to issue a credential in a compatible format. & depends($v$, $i$, NFR5) \\
            \rowcolor{gray!10}
            & & $i$ & $i$ & The data owner depends on the issuer for a credential to be usable. & depends($o$, $i$, NFR5) \\

            \bottomrule
        \end{tabularx}
        \vspace{3px}
        \begin{tablenotes}
            \item[a] Primary Responsibility
            \item[b] Ownership
        \end{tablenotes}
    \end{threeparttable}
\end{table*}

%% file: sections/5_conclusions.tex
\section{Conclusions and Future Work}\label{sec:conclusions}

    This work addresses the limited integration of the individual user perspective into the requirements engineering process in the DI/SSI domain. 
    While prior work consolidated SSI principles, which comprehensively represent quality requirements or NFR of a DI/SSI system, they did not explicitly embed user roles, responsibilities, and dependencies. 
    Building on existing requirements, this work refined and decomposed SSI principles into 24 atomic NFR and systematically mapped them to DI/SSI design patterns, as well as participating users, namely data owner, issuer, verifier, and system. 
    The results highlight the central role of the data owner, who bears primary responsibility for most NFR, thereby supporting the user-centric foundations of DI/SSI. 
    At the same time, the allocation of responsibilities and ownership clarified how issuers, verifiers, and the system contribute to the fulfillment of security, privacy, interoperability, and usability objectives. 
    Furthermore, the proposed dependency model formalizes relationships among actors by explicitly specifying how actors rely on one another to realize requirements. 
    As a result, the model captures functional dependencies of DI/SSI systems.
    
    Future work will focus on identifying the capabilities of each actor in DI/SSI systems, building on the introduced dependency model.  
    Followed by systematic operationalization of NFR, derivation of FR, and development of a functional model for DI/SSI systems. 
    Additionally, mapping design patterns to NFR revealed that not all NFR are addressed by the patterns. 
    Identifying and describing missing design patterns would enable a more comprehensive and complete overview of the DI/SSI system architectures.